\documentclass[11pt,a4paper]{article}
\usepackage{jcappub}
\usepackage{graphicx,amssymb}
\usepackage{graphicx}
\usepackage{url}

\def\be{\begin{equation}}
\def\ee{\end{equation}}
\def\ba{\begin{align}}
\def\ea{\end{align}}

\def\ap{\approx}

\def\lsim{\raise0.3ex\hbox{$\;<$\kern-0.75em\raise-1.1ex\hbox{$\sim\;$}}}
\def\gsim{\raise0.3ex\hbox{$\;>$\kern-0.75em\raise-1.1ex\hbox{$\sim\;$}}}

\def\theta{\vartheta}

\renewcommand{\vec}[1]{\boldsymbol{#1}}

\title{Reconciling cosmic ray diffusion with Galactic magnetic field models}

\author[a]{G.~Giacinti,}
\author[b]{M.~Kachelrie\ss,}
\author[c,d]{D.~V.~Semikoz}

\affiliation[a]{Max-Planck-Institut f\"ur Kernphysik, Heidelberg, Germany}
\affiliation[b]{Institutt for fysikk, NTNU, Trondheim, Norway}
\affiliation[c]{APC, Universite Paris Diderot, CNRS/IN2P3, CEA/IRFU,
Observatoire de Paris, Sorbonne Paris Cite, 119 75205 Paris, France}
\affiliation[d]{National Research Nuclear University MEPHI (Moscow Engineering 
Physics Institute), Kashirskoe highway 31, 115409 Moscow, Russia}

\abstract{
We calculate the diffusion coefficients of charged cosmic rays (CR) 
propagating in regular and turbulent magnetic fields. If the magnetic field 
is dominated by an isotropic turbulent component, we find that CRs reside too 
long in the Galactic disc. As a result, CRs overproduce secondary nuclei like 
boron for any reasonable values of the strength and the coherence length of 
an isotropic turbulent field. We conclude therefore that the propagation of 
Galactic CRs has to be strongly anisotropic because of a sufficiently strong 
regular field and/or of an anisotropy in the turbulent field. As a 
consequence, the number of sources contributing to the local CR flux is 
reduced by a factor ${\cal O}(100)$ compared to the case of isotropic CR 
diffusion.
}

\date{\today}


\begin{document}

\maketitle

\section{Introduction}

The propagation of Galactic cosmic rays (CR) is usually described by 
empirical diffusion models where the energy dependence of the diffusion 
coefficient $D(E)=D_0(E/E_0)^\beta$ is obtained fitting 
observations~\cite{Ginzburg:1990sk,SMP07}. In such models, discrete CR sources
are approximated by a continuous distribution filling a disk with typical
vertical height $h_{\rm d} \sim 0.2$ kpc, while CRs diffuse in a CR halo 
of much larger height $h \sim 3-5$\,kpc. Cosmic rays either interact or
escape in the intergalactic space when they reach the boundary of the 
CR halo. The solutions to these diffusion models are derived typically
in the steady-state regime, i.e.\ any time-dependence in the injection
history of CRs is neglected.

An important constraint on these models comes from ratios of stable primaries 
and secondaries produced by CR interactions on gas in the Galactic disk, 
which depend on the grammage $X$ crossed by CRs. In particular, the 
boron-to-carbon (B/C) ratio measured by the AMS-02 experiment has been 
interpreted by the
collaboration as being consistent with a $\beta\simeq 1/3$ power law  
for the energy dependence of the diffusion coefficient~\cite{AMS-BC}. 
The normalisation $D_0$ is only weakly constrained using measurements
of stable nuclei, since the grammage scales as 
$X\propto (H/h) \times h^2/D = hH/D$ with the height of the CR halo  $H$ and
the height of the matter disc $h$.
Therefore a larger value of $D_0$ can be compensated by an increase of the 
CR halo size $H$. This degeneracy can be broken considering the ratio of 
radioactive isotopes as, e.g., $^{10}$Be/$^{9}$Be: Fitting successfully
these ratios requires a relatively large CR halo, $H\simeq 5$\,kpc,
which in turn leads to relatively large values of the normalisation
constant $D_0=(3-8)\times 10^{28}$cm$^2$/s at $E_0=10$\,GeV~\cite{fit}.

At the fundamental level, charged particles scatter on inhomogeneities
of the Galactic magnetic field (GMF). For a given GMF model, the 
trajectories of charged particles can be calculated and the diffusion 
coefficients can be determined  
numerically~\cite{diff1,diff2,diff3,trans,fila1,fila2}.
The authors of the recent study Ref.~\cite{Subedi:2016xwd} devised a 
theory allowing to calculate analytically the diffusion coefficient 
in different energy ranges, for isotropic turbulence with no 
regular field, and they confronted it with numerical simulations. 
Since calculations are computationally
expensive, they are usually restricted to energies
above $10^{14}$\,eV. On the other hand, empirical diffusion models
are applied mainly below $10^{13}$\,eV. This
missing overlap in energy may be the reason why---except for a few 
cases---a confrontation of the two methods and their conclusions 
seems to be missing in the literature. 
We noted earlier in Ref.~\cite{fila1} that the diffusion coefficient
calculated numerically in a pure random field with $B_{\rm rms}=4\,\mu$G
is smaller than the one extrapolated in the diffusion picture from lower 
energies. In Refs.~\cite{escape1,escape2}, we showed that the grammage
crossed by CRs in the original Jansson-Farrar model~\cite{JF} for the GMF
is a factor of a few to 10 too large.
The aim of the present work is to investigate these discrepancies in
more detail. In particular, we want to determine qualitatively 
those properties of a GMF model which determine if the resulting CR 
propagation can be reconciled with the results obtained in diffusion 
models.  We examine therefore several toy models which allow us to 
isolate some key properties a successful GMF model should possess.

This article is structured as follows: We start summarising in 
Sec.~\ref{approach} how we model  magnetic fields, propagate CRs 
and calculate from their trajectories the diffusion tensor. 
Then we investigate in Sec.~\ref{models} various toy models and 
derive the corresponding diffusion coefficients. Finally, we 
discuss the impact of anisotropic diffusion on Galactic CR physics
and summarise additional evidence disfavouring the traditional steady-state
picture in Sec.~\ref{dis} before we conclude.

\section{Theoretical framework}             \label{approach}

The GMF consists of a regular  component which is ordered on kpc
scales and of a turbulent field. On sufficiently small scales, 
the regular part can be approximated as a uniform field and we will
therefore not consider specific GMF models.
The turbulent component is usually modelled as a Gaussian random
field; such a field is fully characterised by its power spectrum 
$\mathcal{P}(\vec k)$. Several theoretical models, such as that of 
Ref.~\cite{Goldreich:1994zz} for incompressible Alfv\'enic turbulence, 
suggest that the turbulence can be 
anisotropic. We consider therefore both isotropic and anisotropic 
turbulence. We assume the turbulence to be static, and consider  
power laws $\mathcal{P}(k)\propto \vec N k^{-\alpha}$ for its power spectrum. 
The first assumption is well 
justified for the purpose of the present paper: Both 
the Alfv\'en velocity $v_{\rm A}$ and the velocity of the 
interstellar medium $u$ are of the order of tens of km/s, which only induces 
negligible changes on the length and time scales considered here. Motivated by
the consistency of the slope of the B/C ratio with Kolmogorov turbulence, 
we choose $\alpha = 5/3$ as index of the power spectrum, i.e.\ we set 
$\mathcal{P}(k)\propto k^{-5/3}$.

The fluctuations in the turbulent magnetic field extend over a
large range of scales, from the dissipation scale $L_{\min}\sim 1$\,AU to 
the outer scale $L_{\max}$, which varies from $L_{\max}\sim 10$\,pc in
the disc to $L_{\max}\sim 150$\,pc in the halo~\cite{coh}. We note that, for 
Kolmogorov turbulence, the coherence length $L_{\rm coh}$ and the outer 
scale $L_{\max}$ are connected by $L_{\rm coh}=L_{\max}/5$.
In order to cope with this large range of length scales, we construct the 
magnetic field in our numerical simulations either on nested grids as 
described in Ref.~\cite{trans}
or as the sum of circularly polarised plane waves following the
approach of Ref.~\cite{GJ}. Both methods allow us to choose the 
effective minimum scale $L'_{\min}$ of fluctuations in the field 
sufficiently small compared with the Larmor radius 
$R_{\rm L,low}=cp_{\rm low}/(eB)$ of the CRs with the lowest 
energy, $cp_{\rm low}$, in each set of calculations. We always choose 
$L'_{\min}$ such that $L'_{\min} \lesssim R_{\rm L,low}/1.3$. 
For example, we use a grid with $256^3$ vertices for the calculations of 
Fig.~\ref{fig:Dreg+turb}, which corresponds to a dynamical range of 
$L_{\max}/L_{\min} = 128$. In this Figure, 
$L_{\min} = (100\,{\rm pc})/128 = 0.78$\,pc, which is 
smaller than the CR Larmor radius at the lowest energy 
considered there, $R_{\rm L,low}(cp_{\rm low}=1\,{\rm PeV})=1$\,pc in a $1\,\mu$G field.

For the calculation of the  diffusion tensor, we propagate individual CRs 
in a prescribed magnetic field (regular and/or turbulent) solving the Lorentz 
force equation. 
We perform the numerical simulations with the code described in 
Refs.~\cite{nuc1,nuc2,trans}.
Having obtained the trajectories $x_i^{(a)}(t)$ of $N$
CRs with energy $E$, we calculate the diffusion tensor as
\be \label{def}
D_{ij}(E)= \lim_{t\to\infty} \frac{1}{2Nt}\sum_{a=1}^N x_i^{(a)}x_j^{(a)} .
\ee
For sufficiently large times $t$, such that CRs propagate
over several coherence lengths, the RHS becomes time-independent.
Diagonalising then the tensor $D_{ij}$, it can be written as
 $D_{ij}={\rm diag}\{ D_\perp, D_\perp, D_\|\}$ where $D_\|$ denotes
the component aligned with the large-scale field.

\section{Diffusion tensor and magnetic field models}
\label{models}

\subsection{Pure isotropic turbulent field}
\label{Iso_Turb}

It is often assumed that the turbulent component of the GMF dominates over 
the regular one, $B_{\rm rms}\gg B_0$. In a first approach, we set therefore
the regular field  $B_0$ to zero and consider a purely turbulent magnetic 
field. In the diffusion picture, one can model the propagation of CRs as a 
random walk with an energy dependent effective step size. For a
pure isotropic random field, one expects therefore  as functional dependence
of the diffusion coefficient
\be  \label{diff}
  D = \frac{cL_0}{3} 
 \left[ (R_{\rm L}/L_0)^{2-\alpha} + (R_{\rm L}/L_0)^2 \right] \,,
\ee
where the condition $R_{\rm L}(E)=L_0$ determines the transition from 
small-angle scattering with $D(E)\propto E^2$ to
large-angle scattering with $D(E)\propto E^{2-\alpha}$. At even higher
energies, CRs enter the ballistic regime and the concept of a
diffusion coefficient becomes ill-defined.

\begin{figure}
\includegraphics[width=0.65\columnwidth,angle=270]{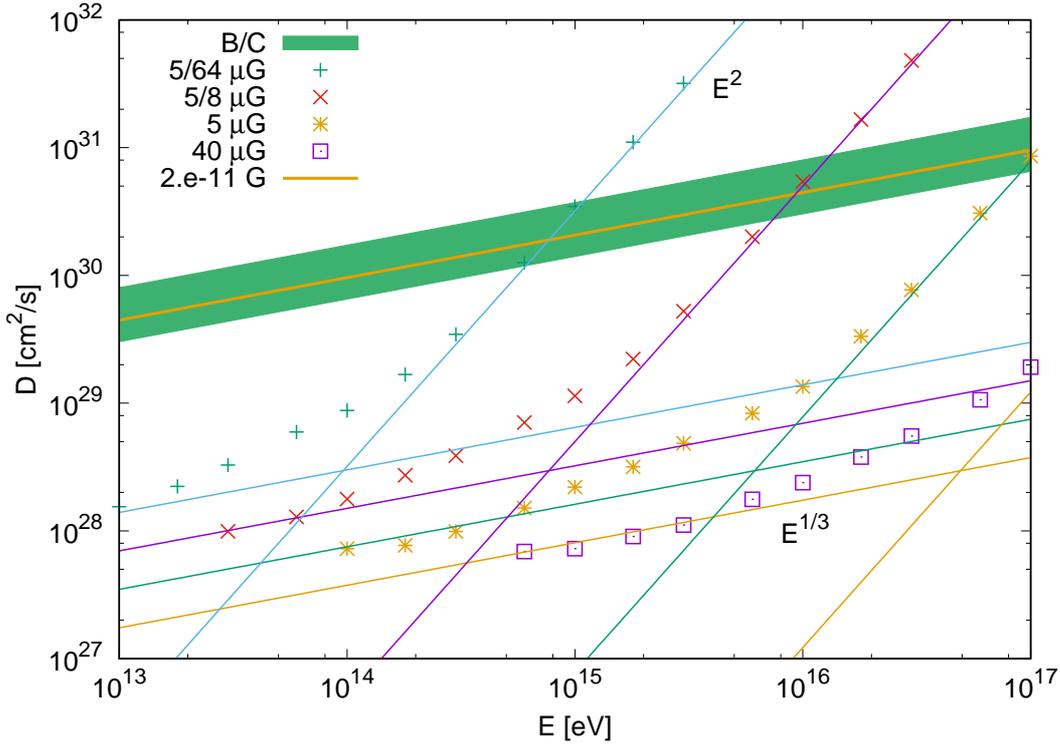}
\caption{CR diffusion coefficient in pure isotropic Kolmogorov turbulence 
with $L_{\max}=25$\,pc and for four values of the turbulence root-mean-square strength: 
$B_{\rm rms} = 40\,\mu$G, $5\,\mu$G, $(5/8)\,\mu$G, 
and $(5/64)\,\mu$G ---see key for symbols. The asymptotic behaviours at 
low and high energies are shown with the solid lines. The lines $\propto E^{1/3}$ differ by a factor $8^{1/3}=2$ in normalization. 
We also plot, in orange, the extrapolated asymptotic low-energy behaviour 
for $B_{\rm rms} = 2 \times 10^{-5}\,\mu$G. It lies within the range of magnetic field strengths 
that satisfy $D_0=(3-8)\times 10^{28}$cm$^2$/s at $E_0=10$\,GeV, 
assuming a \lq\lq Kolmogorov\rq\rq\/ extrapolation to high energies (B/C constraints 
from Refs.~\cite{Ginzburg:1990sk,SMP07}, green area).
\label{iso1}}
\end{figure}

In Fig.~\ref{iso1}, we show the diffusion coefficients numerically
calculated using Eq.~(\ref{def})
for three different magnetic field strengths $B_{\rm rms}$.
The turbulent field was modelled as a homogeneous, isotropic Gaussian 
random field following a Kolmogorov power-spectrum 
$\mathcal{P}(k)\propto k^{-5/3}$ with $L_{\max}=25$\,pc. 
We note first that the scaling of the diffusion coefficient with energy 
and magnetic field strength expected from Eq.~(\ref{diff}) is numerically
reproduced. In particular, the two regimes, i.e.\ the transition from 
$\beta=1/3$ to $\beta=2$ can be clearly
seen. The transition scale $L_0$ should scale with the coherence length 
as $L_0\propto L_{\rm coh}$, but the proportionality factor has to be determined 
numerically. We find that $L_0 \simeq L_{\rm coh}/(2\pi)$ provides a good 
fit to our numerical results, which agrees with the findings of 
Ref.~\cite{diff1}. Having determined $L_0$,  we can extrapolate $D(E)$ to 
low energies and/or large coherence lengths, and compare it to the values 
found by fitting secondary ratios like B/C. 

\begin{figure}
\begin{center}
\includegraphics[width=0.8\columnwidth]{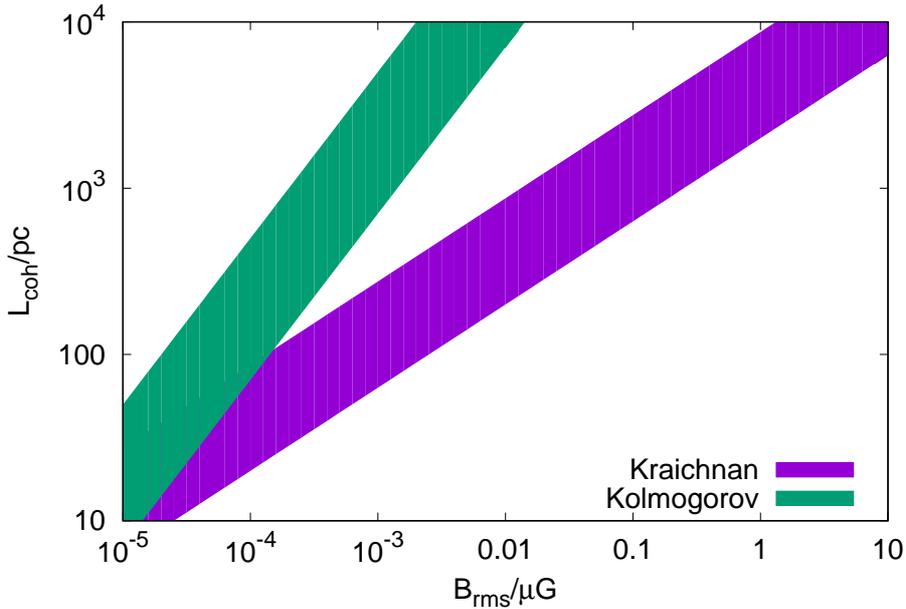}
\end{center}
\caption{Allowed ranges of $B_{\rm rms}$ and $L_{\rm coh}$ compatible
with $D_0=(3-8)\times 10^{28}$cm$^2$/s at $E_0=10$\,GeV for
Kolmogorov and Kraichnan turbulence. These ranges should be compared with 
the typical order-of-magnitude values that are relevant for the Galactic 
magnetic field: $B_{\rm rms} \sim (1 - 10)\,\mu$G and 
$L_{\rm coh} \lesssim$~a few tens of pc.
\label{iso2}}
\end{figure}

Requiring that the diffusion coefficient lies in the range 
$D=D_0=(3-8)\times 10^{28}$cm$^2$/s at $E_0=10$\,GeV, we can determine the
possible range of magnetic field strengths and coherence lengths. 
The allowed ranges of 
$B_{\rm rms}$ and $L_{\rm coh}$ are shown in Fig.~\ref{iso2}, both for
Kolmogorov and Kraichnan turbulence. The weak energy-dependence of 
$D(E)\propto E^{1/3}$ (for Kolmogorov turbulence) requires a
reduction of $B_{\rm rms}$ by a factor $\sim 1/(100)^3 =10^{-6}$, when 
keeping the coherence length $L_{\rm coh}$ fixed. Insisting instead on 
$B_{\rm rms}\sim {\rm few}\times \mu$G, the coherence length $L_{\rm coh}$ 
would have to be comparable to the size of the Galactic halo.

These results demonstrate that the propagation of Galactic CRs cannot be
described as isotropic diffusion: For reasonable values of the
strength and coherence length of a dominantly turbulent isotropic field, 
CRs reside too long in the Galactic disc, overproducing therefore 
secondary nuclei like boron. This problem can be avoided, if the regular 
field is sufficiently strong, such that the diffusion parallel to this regular 
field becomes considerably faster than in the isotropic diffusion picture. 
Additionally, the turbulence itself may be anisotropic, which can further 
increase the parallel diffusion coefficient of CRs. Then the strongly 
anisotropic propagation of CRs 
leads to a more effective escape of CRs from the Galaxy.

\subsection{Isotropic turbulence with a uniform regular field}
\label{Turb_w_Reg}

\begin{figure}
\includegraphics[width=\columnwidth]{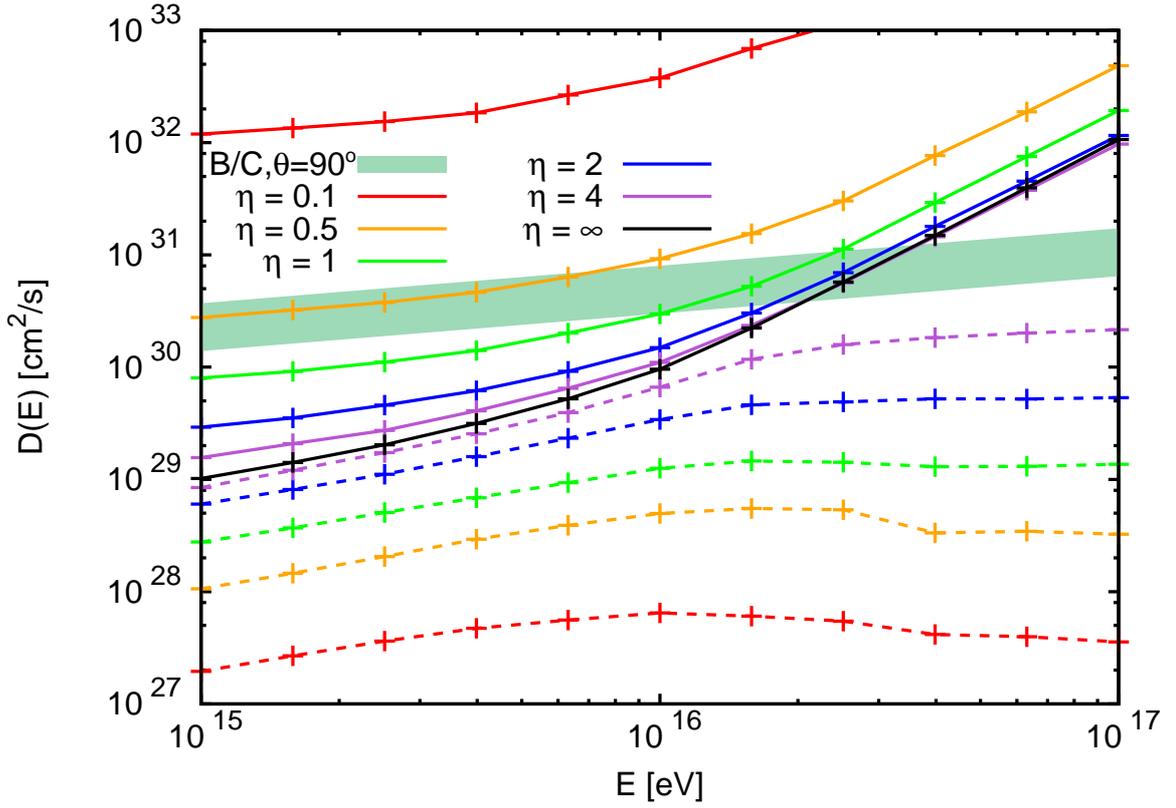}
\caption{
Parallel (solid lines) and perpendicular (dashed lines) diffusion coefficients for isotropic Kolmogorov turbulence with a superimposed regular field. Results are presented for six levels of turbulence: $\eta = 0.1$ (red lines), 0.5 (orange), 1 (green), 2 (blue), 4 (purple), $\infty$ (i.e. pure turbulence, black). Total magnetic field strength set to $B_{\rm tot} = \sqrt{B_{\rm rms}^{2} + B_0^{2}} = 1\,\mu$G, and outer scale of the turbulence equal to $L_{\max} = 100$\,pc. Green area for the \lq\lq Kolmogorov\rq\rq\/ extrapolation to high energies of the value $D_0=(3-8)\times 10^{28}$cm$^2$/s inferred at $E_0=10$\,GeV from the B/C ratio~\cite{Ginzburg:1990sk,SMP07}, and relevant for the limiting case $\theta = 90^{\circ}$. See Fig.~\ref{fig:Xeta} (left panel) for other values of $\theta$.
\label{fig:Dreg+turb}}
\end{figure}

We add next a uniform magnetic field directed along the $z$ direction
to the isotropic turbulent field. As a result, the propagation of CRs 
becomes anisotropic, leading to a diffusion tensor with 
$D_{ij}={\rm diag}\{ D_\perp,D_\perp,D_\| \}$ and $D_\|>D_\perp$.
Figure~\ref{fig:Dreg+turb} shows $D_\|$ (solid lines) and $D_\perp$ (dashed lines) for six values of the 
level of turbulence $\eta = B_{\rm rms}/B_0$: $\eta = 0.1$ (red lines), 0.5 (orange), 
1 (green), 2 (blue), 4 (purple), and $\infty$. The limit $\eta\to\infty$ 
corresponds to isotropic turbulence, and thence isotropic diffusion. 
The total magnetic field strength is chosen as 
$B_{\rm tot} = \sqrt{B_{\rm rms}^{2} + B_0^{2}} = 1\,\mu$G, and the outer scale of the turbulence 
is set to $L_{\max} = 100$\,pc. Decreasing 
$\eta$, the difference between $D_\|$ and $D_\perp$ increases, while keeping the 
order $D_\|>D_\infty>D_\perp$ intact, where $D_\infty(E)$ denotes the diffusion 
coefficient for pure isotropic turbulence. 
We note, as previously pointed out by~\cite{diff3,Snodin:2015fza}, that the 
slope of $D_\perp$ is larger than 1/3---it is close to 1/2. Assessing whether this slope 
continues down to GeV energies or not is difficult due to the limited energy 
range one can probe. If it does, this would only make 
perpendicular diffusion even less relevant for the escape of CRs from the
Galaxy.

Next we want to connect these results with measurements such as the B/C ratio
which constrain the grammage crossed by CRs before escape. The 
level of turbulence $\eta$ that is required to satisfy contraints from the grammage 
depends on the geometry of the regular Galactic magnetic field. We note that $D_\perp < D_\infty$, and that 
the perpendicular diffusion coefficient is therefore too small at any value of $\eta$ to accommodate for a 
perpendicular escape of CRs in the halo, with a purely toroidal regular Galactic magnetic field.

For the sake of the argument, let us start with the simple, limiting case where the regular field is perpendicular to the 
Galactic disc, and where CRs escape at a distance $H$ in the halo, along this field. In this case, the 
constraints that are used in Sect.~\ref{Iso_Turb} for $D_\infty$ can now be applied to $D_\|$. The green area in
Fig.~\ref{fig:Dreg+turb} corresponds to the \lq\lq Kolmogorov\rq\rq\/ extrapolation 
($D(E) \propto E^{1/3}$) of the value $D_0=(3-8)\times 10^{28}$\,cm$^2$/s, inferred from the B/C ratio
at $E_0=10$\,GeV by the authors of Refs.~\cite{Ginzburg:1990sk,SMP07}. 
One can see in Fig.~\ref{fig:Dreg+turb} that cases with $\eta\geq 1$ all give
too small parallel diffusion coefficients to accommodate for
the value required by the boron-to-carbon ratio. On the contrary, the parallel diffusion
coefficients for cases with a strong regular magnetic field ($\eta < 1$) 
are able to fulfil the constraints from the B/C ratio. 
The solid orange line for $\eta = 0.5$ is at the required level, assuming
Kolmogorov turbulence and $B_{\rm tot} = 1\,\mu$G. A slightly lower value of
$\eta$ would be preferred for $B_{\rm tot} \gtrsim 4\,\mu$G. This good match 
between the diffusion coefficient required from the B/C ratio and $D_\|$ for
$\eta = 0.5$ assumes however that the regular field in the halo is
perpendicular to the Galactic plane. In reality, this field is more likely to
be at an angle $\theta$ to the plane, such as e.g.\ for the
\lq\lq X-field\rq\rq\/
in the GMF model of Ref.~\cite{JF}. If so, CRs have to travel longer distances
along the regular field before escape, and therefore lower values of $\eta$
would be required.

\begin{figure}
  \includegraphics[width=0.45\columnwidth]{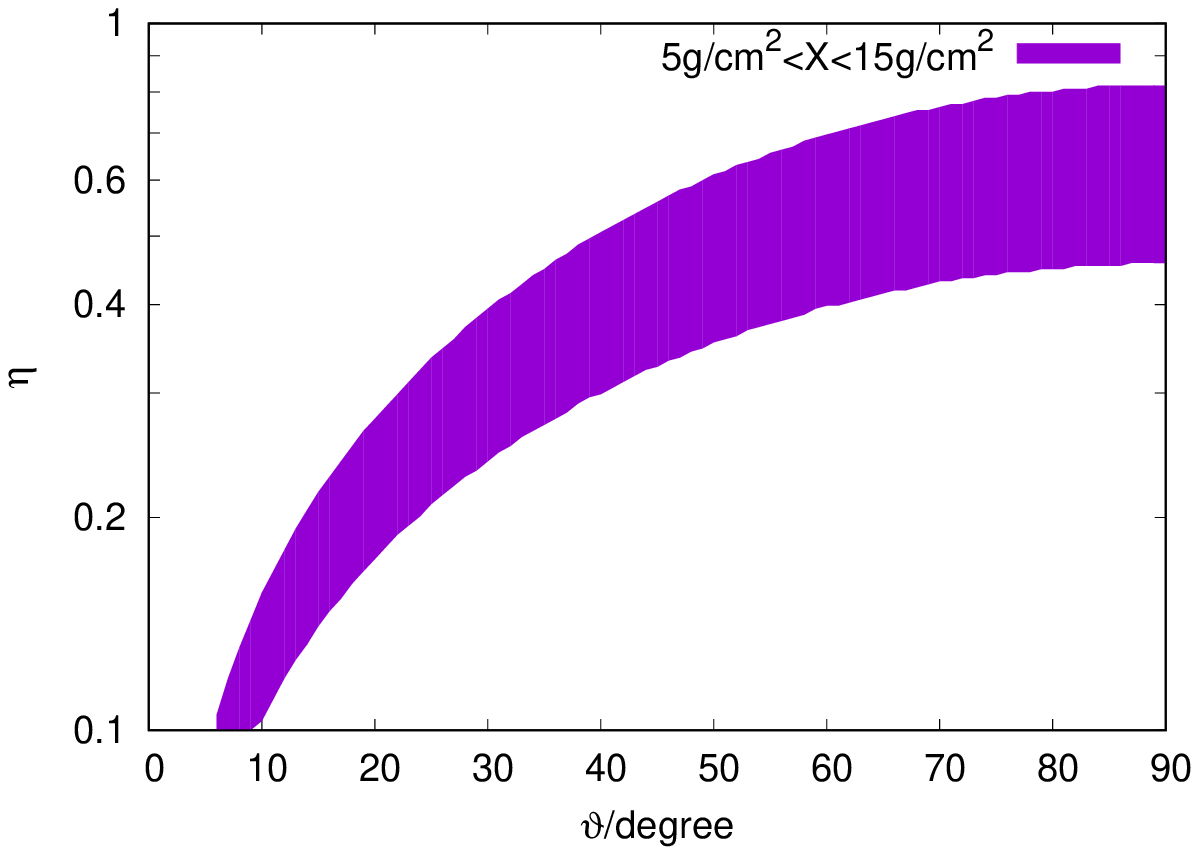}
  \includegraphics[width=0.45\columnwidth]{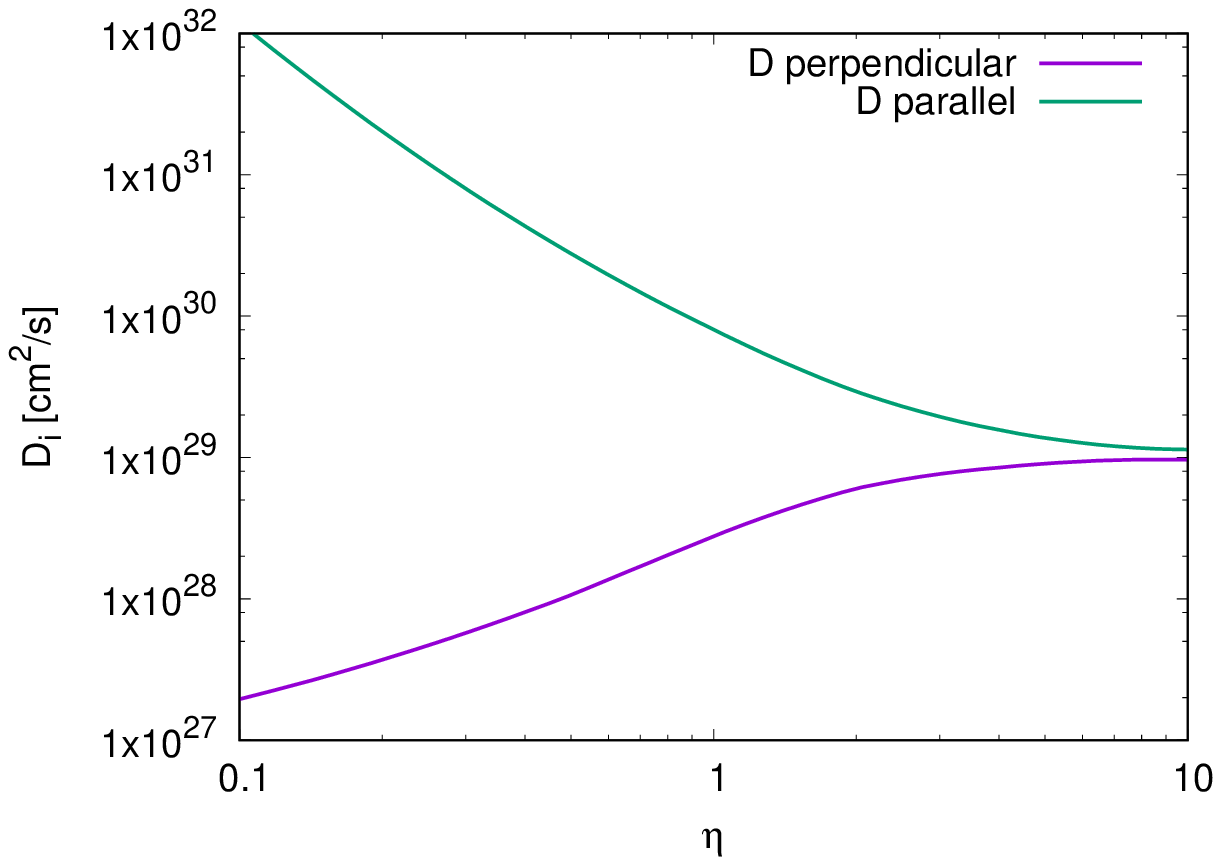}
  \caption{
    Left: Grammage $X$ crossed by CRs in a \lq\lq disc and halo\rq\rq\/ model, for the diffusion coefficients shown in Fig.~\ref{fig:Dreg+turb}, as a function of the tilt angle $\theta$ between the regular magnetic field and the Galactic plane, and of the turbulence level $\eta$.
    Right: Fit of the diffusion coefficients $D_\|$ and $D_\perp$ from
    Fig.~\ref{fig:Dreg+turb} at $E=10^{15}$\,eV as a function of $\eta$.
\label{fig:Xeta}}
\end{figure}

In order to constrain this case quantitatively, we consider the following toy
model: We assume a thin matter disc with density $\rho/m_{\rm p}\simeq 1$/cm$^3$
and height $h=150$\,pc around the Galactic plane. Cosmic rays propagate inside
a larger halo of height $H=5$\,kpc. The regular magnetic field inside this
disc and halo has a tilt angle $\theta$ with the Galactic plane, so that the component
of the diffusion tensor relevant for CR escape is given by
\be
D_z = D_\perp\cos^2\theta + D_\|\sin^2\theta \,.
\ee
Using a simple leaky-box model, the grammage follows as
$X = c \rho hH /D_z$.
Using now as allowed region for the grammage e.g.\ $5\leq X\leq 15$\,g/cm$^2$,
the permitted region in the $\theta$--$\eta$ plane shown in the left panel of
Fig.~\ref{fig:Xeta} follows. For not too large values of the tilt angle,
$\theta\lsim 30^\circ$, the regular field should strongly dominate,
$\eta\lsim 0.35$.
Note that the grammage is a rather steep function of $\eta$. While the
exact contour of the allowed region depends on the assumed parameters, the
conclusion that the regular field should dominate, i.e.\ that $\eta$ is
small, does not depend on the exact values of these parameters.

Let us now consider as
an example of concrete GMF model the one of Jansson and Farrar~\cite{JF}.
We showed in Refs.~\cite{escape1,escape2} that one can reproduce the
correct grammage CRs cross, if one reduces the turbulent field in this model by a
factor $\approx 8-10$. We found that this yields an average turbulence level of 
$\eta \simeq 0.25$ along CR trajectories. CRs propagate mainly along the regular field 
and, since the field in this model contains a $z$ component,
CRs propagate efficiently towards the Galactic halo. In this model, 
$\theta \simeq 10.2^{\circ}$ at Earth, and $\theta$ increases further in the halo 
(with $\theta < 49^{\circ}$ at relevant Galactocentric radii). We estimate as
a typical value for CR escape  $\theta \approx 20^{\circ}$. As can 
be seen in Fig.~\ref{fig:Xeta} (left panel), such a value of $\theta$ is 
consistent with $\eta \approx 0.25$, which corresponds to the aforementioned value
of turbulence level determined in Refs.~\cite{escape1,escape2}. 
This agreement shows that our toy model with a tilted, 
uniform field in the halo provides, in a first approximation, an estimate of the 
required turbulence levels for a GMF model with a non-zero magnetic field component 
perpendicular to the Galactic plane.

In the right panel of Fig.~\ref{fig:Xeta}, we show a fit of the diffusion
coefficients $D_\|$ and $D_\perp$ from Fig.~\ref{fig:Dreg+turb} at
$E=10^{15}$\,eV as a function of $\eta$. Using this fit, one may apply the
same analysis as above to other toy models of the GMF geometry: For instance,
one may estimate the resulting grammage in models without a $z$ component of
the regular magnetic field, where CR escape proceeds along the spiral arms
towards the outer regions of the Galactic disc.

\subsection{Anisotropic turbulent field}
\label{Aniso_Turb}

\begin{figure*}
\begin{center}
\includegraphics[width=0.95\columnwidth]{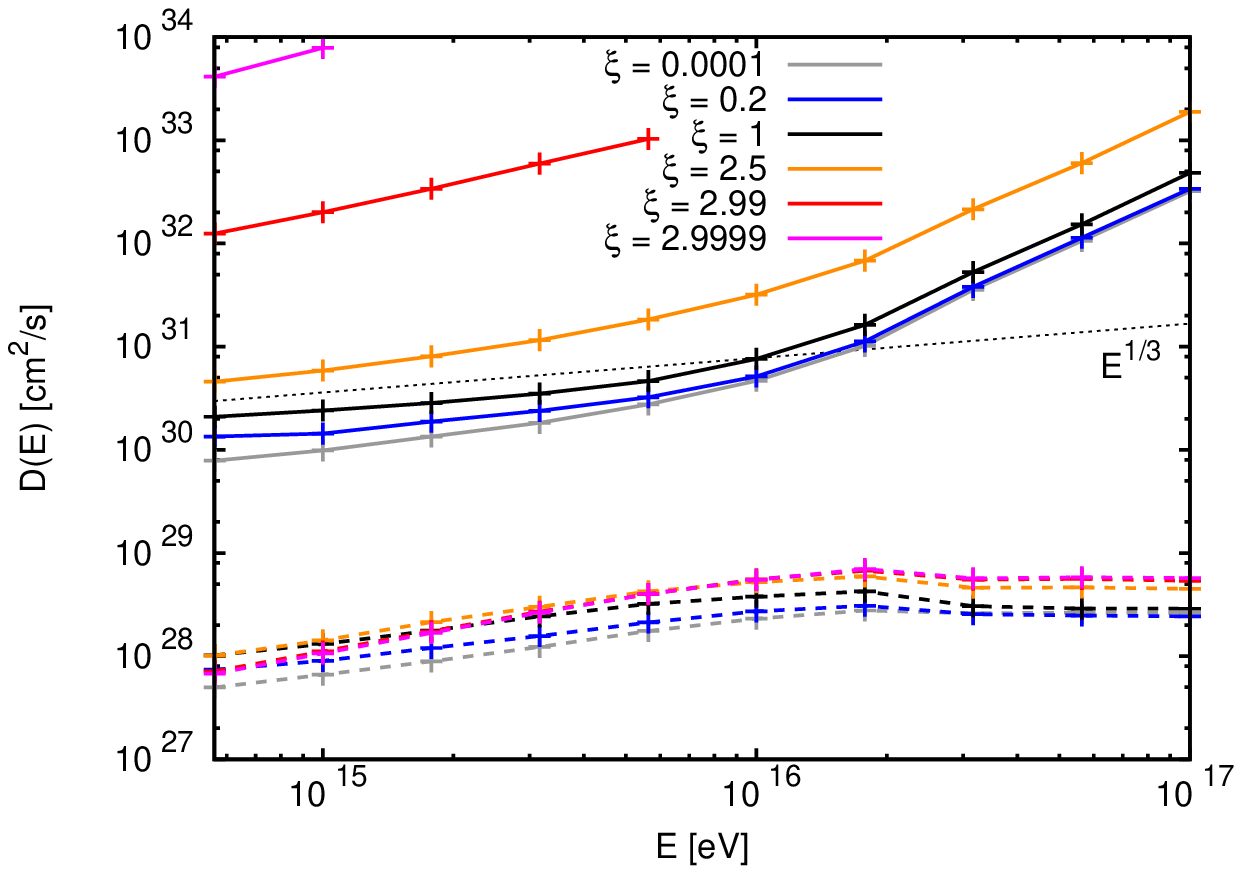}
\includegraphics[width=0.95\columnwidth]{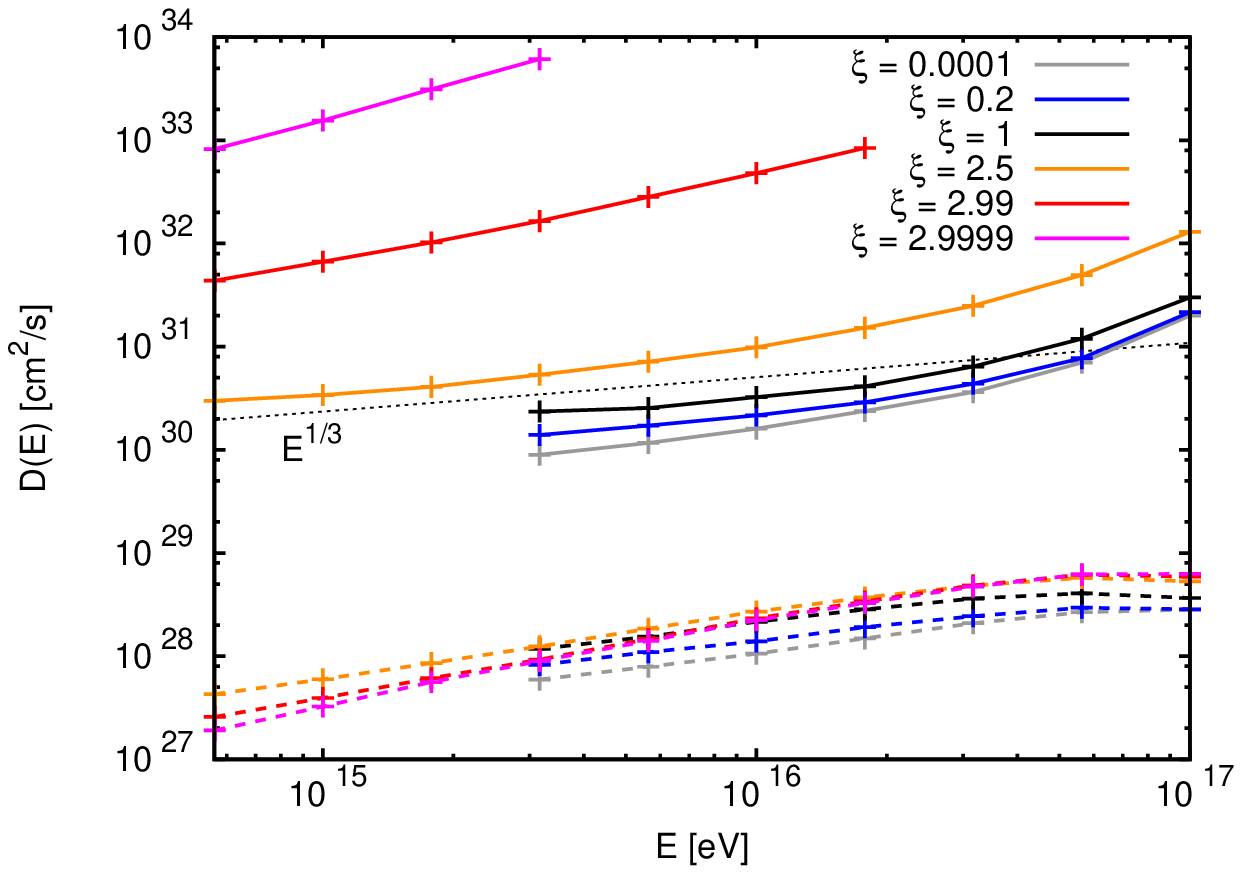}
\caption{Parallel (solid lines) and perpendicular (dashed lines) diffusion coefficients for anisotropic turbulence with $\xi = 0.0001, 0.2, 2.5, 2.99, 2.9999$, and with a superimposed regular field satisfying $\eta = 0.5$ ---see text for the definition of $\xi$. $L_{\max} = 100$\,pc, and $B_{\rm tot} = 1\,\mu$G (upper panel) or $4\,\mu$G (lower panel). Black line for isotropic turbulence ($\xi = 1$). Grey, blue, orange, red, and magenta lines respectively for $\xi = 0.0001$, 0.2, 2.5, 2.99, and 2.9999. To guide the eye, the thin black dotted lines correspond to $D(E) \propto E^{1/3}$ (arbitrary normalizations).}
\label{AnisoTurb_w_Reg}
\end{center}
\end{figure*}

\begin{figure}
  \includegraphics[width=0.49\columnwidth]{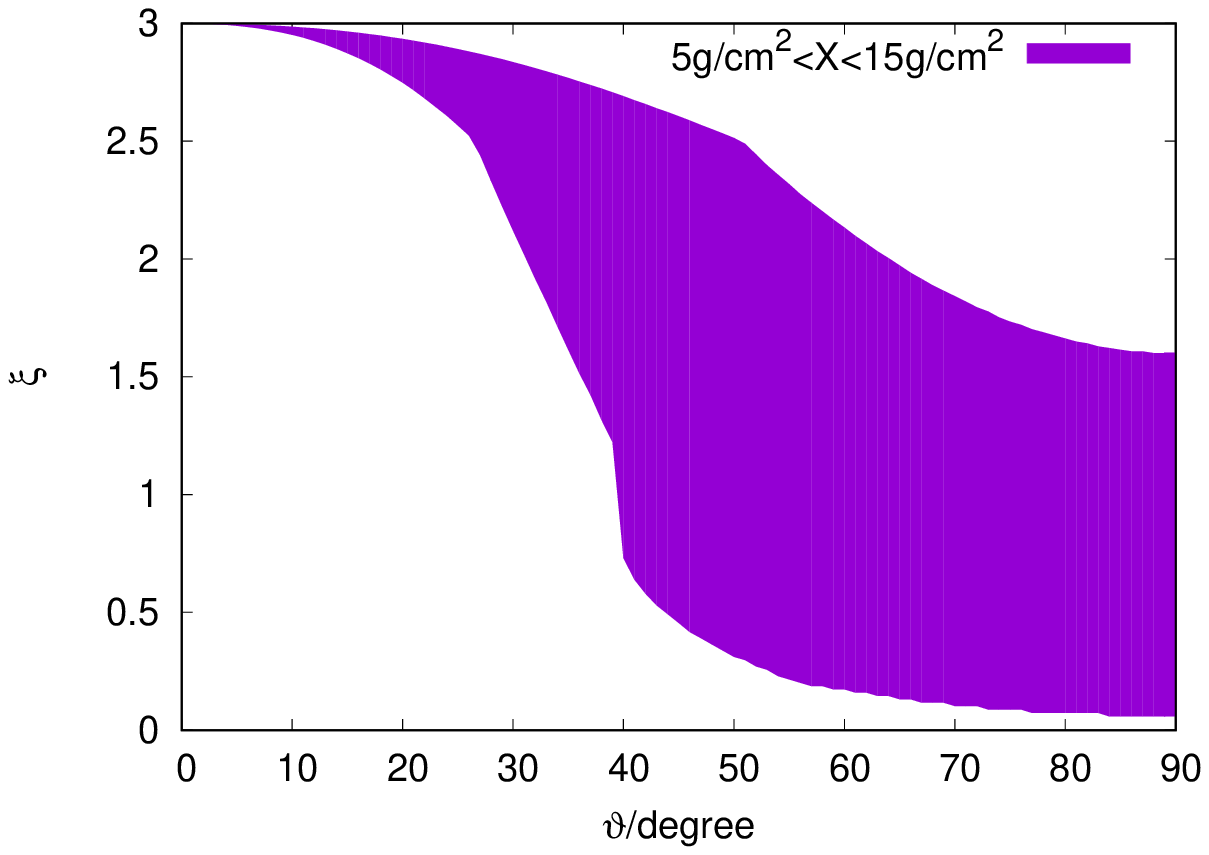}
  \includegraphics[width=0.49\columnwidth]{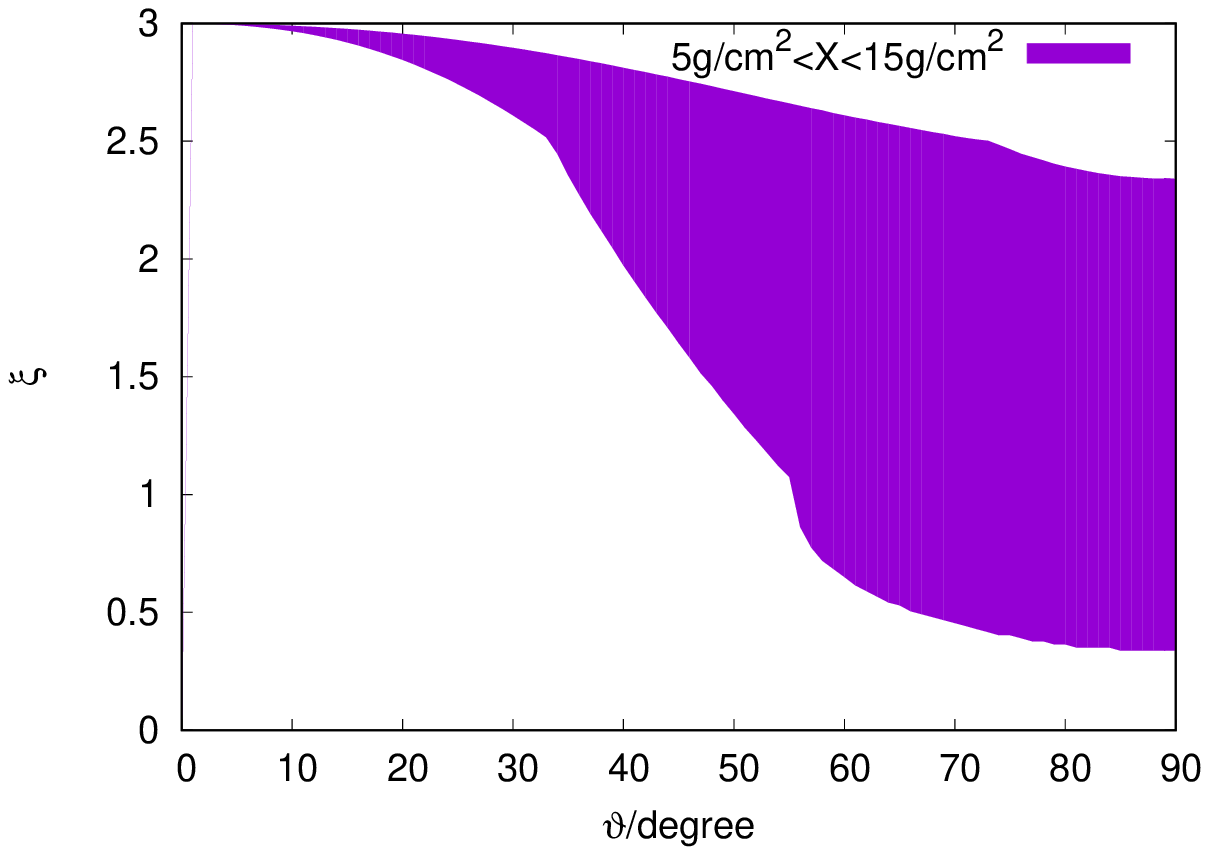}
\caption{Grammage $X$ crossed by CRs in a \lq\lq disc and halo\rq\rq\/ model, for the diffusion coefficients shown in Fig.~\ref{AnisoTurb_w_Reg}, as a function of the tilt angle $\theta$ between the regular magnetic field and the Galactic plane, and of the level of anisotropy $\xi$; left panel for $B_{\rm tot} = 1\,\mu$G, right panel for $B_{\rm tot} = 4\,\mu$G.
\label{fig:Xeta4}}
\end{figure}

As noted above, the presence of a strong regular field such that $\eta\ap 0.3$ 
would solve the discrepancy between estimates of the CR diffusion coefficient 
from the B/C ratio and the calculations presented for pure isotropic 
turbulence ($\eta = \infty$). Another possible solution is that interstellar 
turbulent magnetic fields are anisotropic. We estimate below how much the 
diffusion coefficient would change depending on the level of anisotropy of 
the turbulence.

Let us start with a model where the anisotropic turbulence is generated in the following phenomenological way. Assuming a non-zero regular field along $z$, we use the same grid as in the previous Sections for isotropic turbulence, but rescale its components as: 
$B_z \rightarrow \sqrt{\xi} \, B_z$, and $B_{x,y} \rightarrow \sqrt{(3-\xi)/2} \, B_{x,y}$, with $0 < \xi < 3$. With this prescription, the value of $B_{\rm rms}$ remains unchanged from that of the initial grid, while the component of the turbulent field in the direction of the regular field can be either enhanced or decreased. We have defined here the direction of the anisotropy as that of the regular field and not of local field lines. This is acceptable as long as $\eta$ is small. This method has the advantages of being computationally inexpensive, providing an easy way to generate the turbulence, and giving intuitively clear results. It does not conserve ${\rm div} \,\vec B = 0$ though, but this is unimportant for the message of the present paper. We verified on a few examples that generating anisotropic turbulence with ${\rm div} \, \vec B = 0$ would not change results significantly: To do so, we used the method of Refs.~\cite{GJ}, and introduced an anisotropy in Fourier space, for instance by generating wave vectors $\vec k$ preferentially perpendicular to the regular magnetic field. This is however computationally more expensive.

In Figure~\ref{AnisoTurb_w_Reg}, we present calculations of the parallel (solid lines) and perpendicular (dashed lines) diffusion coefficients with the grid method, for a turbulence level $\eta = 0.5$. The outer scale of the turbulence is set to $L_{\max}=100$\,pc, and the total magnetic field strength $B_{\rm tot} = \sqrt{B_{\rm rms}^{2} + B_0^{2}}$ is equal to $1\,\mu$G in the upper panel, and $4\,\mu$G in the lower one. Five levels of anisotropy are shown: Two where the component $B_{z}$ of the turbulence is reduced, $\xi = 0.0001$ (grey lines) and 0.2 (blue), and three where it is enlarged, $\xi = 2.5$ (orange lines), 2.99 (red) and 2.9999 (magenta). For reference, the case of isotropic turbulence ($\xi = 1$) is shown with the solid and dashed black lines. The thin black dotted lines in each panel correspond to $D(E) \propto E^{1/3}$, and have arbitrary normalizations. They are plotted to guide the eye, and show 1/3 slopes.

As can be seen in Fig.~\ref{AnisoTurb_w_Reg}, the value of the perpendicular diffusion coefficient does not vary significantly with $\xi$. On the contrary, the variations in the parallel diffusion coefficient can be substantially larger for some values of $\xi$. For the two cases where the turbulence is enhanced in the directions perpendicular to the regular field ($\xi = 0.0001$ and $\xi = 0.2$), both the parallel and perpendicular diffusion coefficients are reduced by a small factor, in comparison with the case of isotropic turbulence. In both panels, this reduction is quite small: no more than a factor $\simeq 2-3$, even in the extreme case of a nearly pure perpendicular geometry of the turbulence. This kind of anisotropy neither helps in increasing the diffusion coefficient, nor has a substantial impact on it.

On the contrary, for $\xi > 1$, the parallel diffusion coefficient increases in comparison with isotropic turbulence. For $\xi=2.5$, it still has not fully reached its asymptotic $\propto E^{1/3}$ behaviour for $B_{\rm tot} = 1\,\mu$G at the lowest energy we consider here, $10^{15}$\,eV. However, this would happen at lower energies. Indeed, for the stronger field shown in the lower panel, the solid orange line reaches a $\propto E^{1/3}$ behaviour below $\simeq 3$\,PeV as can be seen by comparing its slope with that of the thin black dotted line. The slope of the solid black line ($\xi = 1$) converges towards $1/3$ more quickly. It is reached below a few PeV even in the upper panel. 
From the lower panel, one can deduce that the  normalisation of the diffusion 
coefficient at low energies for $\xi=2.5$ is larger than that for $\xi=1$ by 
a factor $\approx 2.7$. 
Such a value is however still small compared with the increase one would need in order to make the cases with $\eta \geq 2$ satisfy the constraints from the grammage, even though $\xi=2.5$ already corresponds to a non-negligible anisotropy. We verified that even in the case of $\xi = 2.9$, $D_\|$ is only increased by a factor of a few more compared with $\xi = 2.5$. 
The parallel diffusion coefficient
$D_\|$  starts for $\xi = 2.99$ to be an order of magnitude larger than for $\xi = 2.5$, see solid red lines in Fig.~\ref{AnisoTurb_w_Reg}. The solid magenta lines for $\xi=2.9999$ show that if nearly all the power of the turbulence is along $B_{z}$, then the diffusion coefficient would increase by orders of magnitude. Indeed, there is almost no power in the two components of the turbulent magnetic field that can change the pitch-angle of the CRs, $B_{x}$ and $B_{y}$. We conclude that anisotropic turbulence can alleviate the constraints put on $\eta$ in Sect.~\ref{Turb_w_Reg}, by allowing for larger turbulence levels $\eta$ to satisfy the constraints from the boron-to-carbon ratio. However, a very large level of anisotropy in $\vec k$-space, $\xi$, would be needed for the difference to be substantial.
Consequently, CRs would propagate also in this case strongly anisotropically.

We calculate now the grammage for $\eta = 0.5$, within 
the toy model with a tilted uniform field in the disc and halo described in Sect.~\ref{Turb_w_Reg}. 
In Fig.~\ref{fig:Xeta4}, we plot in magenta the allowed area of $\theta$--$\xi$ parameter space that corresponds to grammages 
$5\leq X\leq 15$\,g/cm$^2$. The left panel is for $B_{\rm tot} = 1\,\mu$G and the right one is for $B_{\rm tot} = 4\,\mu$G. 
One can see from these plots that, for such a turbulence level $\eta$, strongly anisotropic fields 
with $\xi\gsim 2.5$ are required for not too extreme 
tilt angles $\theta \lesssim 35^{\circ}$.

\begin{figure}
\includegraphics[width=0.95\columnwidth]{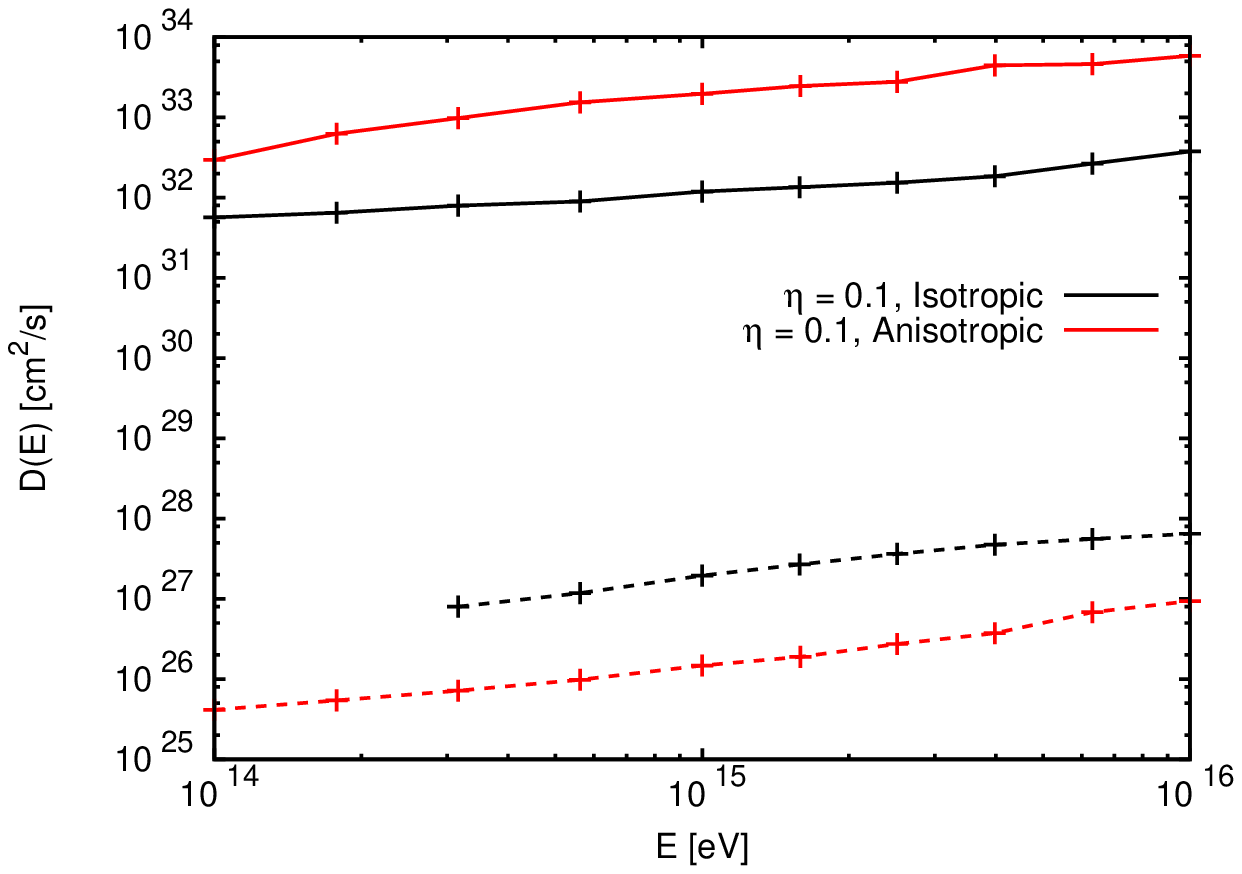}
\caption{Parallel (solid lines) and perpendicular (dashed) diffusion coefficients for isotropic (black) and Goldreich-Sridhar (red, spectrum from Ref.~\cite{Chandran:2000hp}) turbulence, with a superimposed regular magnetic field satisfying $\eta = 0.1$. $L_{\max} = 100$\,pc and $B_{\rm tot} = 1\,\mu$G.}
\label{AnisoTurbGS}
\end{figure}

Finally, we note that power-spectra in the parallel and perpendicular directions 
may be different. For instance, this is the case for Goldreich-Sridhar 
turbulence~\cite{Goldreich:1994zz}, where the power spectrum in the perpendicular direction
is Kolmogorov-like, and that in the parallel direction is $\propto k^{-2}$. While such a type
of turbulence is more isotropic on scales close to 
the outer scale, it becomes increasingly anisotropic at large wave-vectors. 
In Fig.~\ref{AnisoTurbGS}, we present with red lines our numerical calculations of the parallel 
(solid lines) and perpendicular (dashed) diffusion 
coefficients for Goldreich-Sridhar turbulence with $L_{\max} = 100$\,pc, 
and with a superimposed regular magnetic field satisfying $\eta = 0.1$ and $B_{\rm tot}=1\,\mu$G. 
We use the power-spectrum suggested by Ref.~\cite{Chandran:2000hp},
\be
\mathcal{P}({\vec k}) \propto k_\perp^{-10/3} g(k_{\|}L_{\max}^{1/3}/k_\perp^{2/3}),
\ee
with $g(y)=1$  for $|y| < 1$ and $g=0$ otherwise.
For reference, we plot with black lines the results for 
isotropic Kolmogorov turbulence with the same $L_{\max}$, $\eta$ and $B_{\rm tot}$. 
One can see that $D_\|$ for this formulation of 
Goldreich-Sridhar turbulence is about an order of magnitude larger than that 
for isotropic Kolmogorov turbulence. On the contrary, $D_\perp$ is about an order of magnitude 
smaller than that for isotropic turbulence. This is compatible with the expectation that 
CR scattering in Goldreich-Sridhar turbulence must be reduced 
with respect to isotropic turbulence~\cite{Chandran:2000hp}. 
The red and black curves do not coincide at the highest energies shown in 
Fig.~\ref{AnisoTurbGS} because the power-spectrum from Ref.~\cite{Chandran:2000hp} 
still contains some anisotropy even at small $|{\vec k}|$, i.e.\ close to the outer scale. While quasi-linear theory predicts 
that the parallel diffusion coefficient at low CR energies would become flatter or even increase with decreasing 
CR energy (see e.g.\ the red lines in the right panel of Fig.~11 from Ref.~\cite{Giacinti:2016tld}), this effect 
is not observed in the energy range we probe here. The considered energies may be too high 
for this effect to become noticeable, or the unavoidable 
slight mismatch between local magnetic field lines and the direction 
of the anisotropy in numerical simulations may have an impact.
While understanding the exact impact of anisotropic turbulence \`a la Goldreich-Sridhar
on CR propagation clearly requires (and deserves) much more detailed numerical investigations,
we may suggest that also in this case a small value of $\eta$ is required to ensure fast
enough CR escape.

\section{Discussion}                \label{dis}

In the standard picture of CR propagation in the Galaxy, CRs are assumed to
diffuse isotropically in a cylindrical box of height $2H$. This generally
accepted picture implies that a large number of CR sources contribute to
the CR flux detected at Earth, at least up to energies of 100\,TeV,
resulting in a smooth CR flux as a function of time and energy.
Hereafter, we  argue that our results question this. 
Strongly anisotropic diffusion of CRs reduces significantly the number 
of CRs sources contributing to the local CR flux. After illustrating this 
point, we review additional evidence for the breakdown of the steady-state 
picture of continuous CR injection  at relatively low energies, 
$E\sim 0.1$--1\,TeV.

\subsection{Number of sources contributing to the local CR flux}         \label{NbSources}

In order to illustrate the impact of anisotropic diffusion on the assumptions 
underlying Galactic CR physics, we consider as an example the number of 
sources contributing to the local CR flux. More specifically, we discuss 
inspired by Ref.~\cite{local,local2} the question if a single 2--3\,Myr source
can dominate the CR flux in the 10\,TeV energy range. Let us assume that
the source injects instantaneously $10^{50}$\,erg in CRs with the power law
$Q(E)=Q_0 (E/E_0)^{-\alpha}$ and $\alpha\simeq 2.2$. We have shown in 
Refs.~\cite{fila1,fila2} that the diffusion approximation can be applied once CRs 
have reached distances from the source that are greater than a few times $L_{\rm coh}$. 
Observations from Refs.~\cite{Frisch:2012zj,Frisch:2015hfa} suggest that the 
local turbulence appears coherent within $\approx 10$\,pc 
from Earth, which provides an estimate of the local value of $L_{\rm coh}$. 
Therefore, the use of the diffusion approximation is justified for a
local source at the distance of interest here, 100--200\,pc from Earth. 
At a given energy,
the functional behaviour of the observed CR flux from a single source
at the distance $L$ and the age $t$ can be divided into three regimes: 
For $2D t\lsim L^2$, the diffuse flux is exponentially suppressed, while for
intermediate times it is
\be \label{intens}
 I(E)\simeq \frac{c}{4\pi} \frac{Q(E)}{V(t)}
\ee
with $V(t)=\pi^{3/2} D_\perp D_\|^{1/2}t^{3/2}$. When the diffusion front reaches 
the edge of the Galactic CR halo with size $H$, CRs start to escape.
Thus in the third, final time regime, the slope of the CR intensity steepens 
as $I(E)\propto E^{-\alpha-1/3}$ in the case of Kolmogorov turbulence.
Note that for all estimates of the type $2D t\sim L^2$ or  $2D t\sim H^2$, 
the relevant component of the diffusion tensor should be used.

We consider first the standard case of isotropic diffusion. Then at the
reference energy $E_\ast=10$\,TeV, the isotropic diffusion coefficient
$D_{\rm iso}$ satisfying the B/C constraints
equals $D_{\rm iso}(E_\ast)\sim 5\times 10^{29}$cm$^2/$s. Thus the size of the
diffusion front is $L(t)=\sqrt{2Dt}\simeq 2.5$\,kpc for  $t=2$\,Myr. Assuming a 
Galactic CR halo size of $\sim 5$\,kpc, we can use Eq.~(\ref{intens}) to 
estimate the contribution of this single source to the observed CR intensity 
as
$$
 E^{2.8}_\ast I(E_\ast) \simeq 
 200 \,{\rm GeV^{1.8} \: sr^{-1}\:s^{-1}\: m^{-2}} \,.
$$
This corresponds only to 1/100 of the observed CR intensity at this energy.  
Therefore one expects that a large number of sources contribute to the
local CR intensity in the case of isotropic diffusion. As a result, 
features of individual sources like a varying nuclear composition or
maximal energies are washed out, and both the primary and secondary
CR fluxes should be smooth.

Moving on to the case of anisotropic diffusion, we read from
Fig.~\ref{fig:Xeta}
using $\eta=0.25$ and $D_{\rm iso}\simeq 2\times 10^{30}$cm$^2/$s valid at
$E=10^{15}$\,eV that $D_\|\simeq 5 D_{\rm iso}$, while
$D_\perp\simeq D_{\rm iso}/500$.  
Hence the volume $V(t)=\pi^{3/2} D_\perp D_\|^{1/2}t^{3/2}$ is reduced by $500/\sqrt{5}\simeq 200$ compared to the usually considered case of isotropic diffusion.
Thus, a single source
can contribute a fraction of order ${\cal O}(1)$ to the total CR intensity.
Consequently, one expects breaks in the CR primary fluxes as a result
of the varying composition and spectral shape of different sources,
resulting in step-like features in the secondary ratios as discussed
in Ref.~\cite{local2}.
Note also that the volume $V$ containing CRs overlaps only partially with
potential observers located in the thin disc of height $h$: For a non-zero
tilt angle $\theta$, the lenght of the ellipsoid contained in the thin disk
is limited by $\simeq h/\sin\theta\simeq 3h$.

\subsection{Additional evidence against a steady-state CR flux}      

The assumption of a continuous injection of CRs is challenged by other
observations too. First, a CR flux $\propto 1/E^{2.7}$ requires an injection 
spectrum $\propto 1/E^{2.4}$, for a $\propto E^{1/3}$ diffusion coefficient 
with energy dependence $\beta\simeq 1/3$~\cite{AMS-BC}. This slope 
significantly differs from the prediction $1/E^{2-2.2}$ of diffusive shock 
acceleration. Second, recent analyses of the photon flux measured by Fermi-LAT 
have shown a flux following a $1/E^{2.4}$ power law in the central 
Galaxy~\cite{NM2015,Yang2016,Fermi_spectrum}, i.e.\ a slope
which is consistent with shock accelerated protons diffusing in Kolmogorov
turbulence. 
Moreover, the CR flux deduced from photons emitted by local Giant Molecular 
Clouds within 1\,kpc from the Sun shows a $1/E^{2.4}$ behaviour at 
$E<20-40$\,GeV and a variable, space-dependent softer flux at higher 
energies~\cite{NMS2017}. Third, the locally measured CR fluxes of nuclei 
have several breaks. In particular, the spectra show a softening above 
$\simeq 10$\,GV, followed by a hardening above several hundred 
GV~\cite{PAMELA_pHe,AMS02_p,AMS02_He,CREAM,CREAM_nuclei,CREAMIII}.
Finally, the slopes of the spectra $dN/dR$ of different nuclei 
differ~\cite{PAMELA_pHe,AMS02_p,AMS02_He}, although both acceleration 
and diffusion depend only on rigidity $R=cp/Z$. 
A possible interpretation of these measurements is that the usually assumed
steady-state regime holds locally only at low energies~\cite{low}, 
while at higher 
energies the contribution of individual sources becomes important.
This interpretation is natural in the picture of anisotropic diffusion, 
since then the number of sources contributing to the local flux is
strongly reduced.

There are several additional experimental signatures which suggest to 
abandon the steady-state regime.  First, measurements of the CR dipole 
show a plateau of the amplitude in the energy range 1--100\,TeV with 
an approximately constant phase~\cite{aniso}. This is inconsistent with 
simple diffusion  models which predict that the anisotropy grows as 
function of energy as $(E/E_0)^\beta$ with $\beta\simeq 1/3$ for 
Kolmogorov turbulence. Second, the positron and antiproton fluxes have the
same slope as the proton
flux~\cite{PAMELA_pos,AMS02_pos,PAMELA_pbar,AMS02_pbar}, while in the
standard diffusion models they decrease relatively to the primary proton flux  
as $E^{-\beta}$. All these anomalies can be explained by a 2--3\,Myr old
local source dominating the CR flux in this energy 
range~\cite{savchenko,local,local2}. Note also that the B/C ratio
found in Ref.~\cite{local2} is consistent with Kolmogorov turbulence.

\subsection{Possible restrictions}      

We note the following three possible caveats: First,
we have considered throughout this work Gaussian random fields. Turbulent
fields obtained in MHD simulations are however intermittent and it has been
argued that this effect reduces on average the deflections of CRs scattering
on magnetic irregularities~\cite{inter}. This study also suggests that 
intermittency enhances anisotropies in the CR propagation. In order to 
quantify the consequences 
of non-Gaussian, intermittent magnetic field additional numerical studies
should be performed in the future.
Second, we have assumed that streaming of 
CRs around their sources does not have a major impact on our final conclusions. 
We note, however, that low-energy CRs escaping from their sources 
should excite the streaming instability~\cite{Ptuskin:2008zz,Malkov:2012qd,Nava:2016szf}, 
and thereby reduce the CR diffusion coefficient around the sources. For a study 
of the impact on the B/C ratio, see Ref.~\cite{DAngelo:2015cfw}. 
Any impact at the higher energies we considered in the 
above two subsections is questionable and strongly depends on the physical 
conditions around the sources, see the appendix of Ref.~\cite{fila2}.
Third, the results presented here are derived in the limit of negligible CR 
advection in the halo. If a strong Galactic wind were to be present in the 
halo, advection could make the effective size of the \lq\lq halo\rq\rq\/ 
from which CRs are able to come back to the disk smaller than the usual 
estimates ($\sim$~a few kpc). This would in turn reduce the normalisation 
of the relevant range of values that fit the B/C data. While advection
should play a role at sufficiently low energies, there is no universal 
consensus on the presence of breaks, e.g.\ in the
B/C ratio, that would indicate the transition to an advection-dominated 
regime~\cite{AMS-BC}. Reference~\cite{Aloisio:2015rsa} suggests
that an advection-dominated
regime is reached at low energy. However, proof of such a regime at 
$\sim$\,GeV energies would still leave open the question whether 
a large-scale Galactic wind has an important impact on CR observables 
at the higher energies considered here.

\section{Conclusions}

We calculated the diffusion coefficients of charged cosmic rays (CR), 
propagating them in regular and turbulent magnetic fields by solving
the Lorentz equation. We showed that CRs reside too long in the Galactic 
disc, if the magnetic field is dominated by an isotropic turbulent 
component for any reasonable combination of field-strengths $B_{\rm rms}$ and
coherence lengths $L_{\rm coh}$. As a result, CRs scattering on gas in the
Galactic disc overproduce secondary nuclei like boron. 
Therefore propagation of Galactic CRs has to be strongly anisotropic, 
because of a sufficiently strong regular field, and/or an anisotropy of the 
turbulent field. A viable solution in our toy model, 
described in Section~\ref{Turb_w_Reg}, is given by 
a tilted magnetic field with $\theta\simeq 20^\circ$ and $\eta\simeq 0.25$,
resulting in $D_\|/5 \simeq 500 D_\perp\simeq D_{\rm iso}$ for the diffusion
coefficients.
As a consequence, the volume that CRs emitted by a single source occupy at intermediate
times is reduced by a factor ${\cal O}(100)$ compared to the case of 
isotropic CR diffusion. Similarly, the number of sources contributing 
to the local CR flux is reduced and, therefore, single sources can
dominate the CR flux already in the TeV energy range.

\acknowledgments

MK would like to thank the Lorentz Center for hospitality during 
the program \lq\lq A Bayesian View on the Galactic Magnetic Field\rq\rq\/ 
and its participants for fruitful discussions.
We thank the referee, Pasquale Blasi, for his helpful comments and suggestions.


\end{document}